# VOICE CHATBOT FOR HOSPITALITY

SAGINA ATHIKKAL[1] and JOHN JENQ[2]

[1]Department of Computer Science, Montclair State University, NJ USA
athikkals1@montclair.edu
[2]Department of Computer Science, Montclair State University, NJ USA
jenqj@montclair.edu

## ABSTRACT

*Chatbot is a machine with the ability to answer automatically through a conversational interface. A chatbot is considered as one of the most exceptional and promising expressions of human computer interaction. Voice-based chatbots or artificial intelligence (AI) devices transform human-computer bidirectional interactions that allow users to navigate an interactive voice response (IVR) system with their voice generally using natural language. In this paper, we focus on voice based chatbots for mediating interactions between hotels and guests from both the hospitality technology providers' and guests' perspectives. We developed a hotel web application with the capability to receive a voice input. The application was developed with Speech recognition and deep synthesis API for voice to text and text to voice conversion, a closed domain question answering (cdQA) NLP solution was used for query the answer.*

## 1. INTRODUCTION

A chatbot is a programming interface that simulates the conversation or "chatter" of a human being through text or voice interactions. Nowadays, chatbots are available in almost many aspects of technology, such as mobile assistants, customer services, e-commerce, and smart devices. It is a type of software which can help the users by automating their conversations and interact with the customers through the messaging platforms. These chatbot-virtual assistants are found useful to handle simple, look-up tasks in business-to-consumer and business-to-business environments. Chatbot virtual assistants are helpful not only to make use of support staff time but also beneficial in providing a level of customer service when the supporting agents aren't available [1]. Chatbots interpret and process user's words or phrases giving them an instant pre-set answer [2]. The most important aspect of implementing a chatbot is selecting the right natural language processing (NLP) engine. If the user interacts with the bot through voice, for example, then the chatbot requires a speech recognition engine. Similar to regular apps, chatbots also have an application layer, a database, APIs, and Conversational User Interface (CUI)[2]. There are structured and unstructured conversations. Chatbots built for structured conversations are highly scripted, it simplifies programming but restricts the kinds of things that the users can ask. In most B2B environments, chatbots are commonly scripted and used to respond to frequently asked questions or perform simple, repetitive calls to action. In sales, a chatbot may be a quick way for sales reps to get phone numbers. For service departments, it assisting service agents in answering repetitive requests. Generally, once a conversation gets too complex for a chatbot, the call or text window will be transferred to a human service agent.

Chatbots such as ELIZA and PARRY were early attempts at creating programs that could at least temporarily fool a real human being into thinking they were having a conversation with another person. PARRY's effectiveness was benchmarked in the early 1970s using a version of a Turing test; testers only made the correct identification of a human versus a chatbot at a level consistent with making a random guess.

Chatbots have come a long way since then. They are built on artificial intelligence (AI) technologies, including deep learning, natural language processing and machine learning (ML) algorithms, and require massive amounts of data. The more an end user interacts with the bot, the better voice recognition becomes at predicting an appropriate response. We can roughly classify chatbots into three categories: (a) Rule-based, this is the simplest type of chatbots. They require user to make a few selections, such as using drop downs or buttons, to give relevant answers. They are slow but is easy to implement. When many conditions or factors are involved in the knowledge base, this approach may not be the best solution [3]. (b) Intellectually independent chatbots: These chatbots learn from the user's inputs and requests by using Machine Learning. This kind of bots are trained in such a way to understand specific keywords and phrases that triggers bot's reply. They train themselves to understand more and more questions with practice and experience [3]. They spot keywords or phrases and provide predefined answer based on these spotted keywords or phrases. (c). AI-powered chatbots: It combines the best from the rule-based and intellectually independent chatbots. These bots understand free language and make sure they solve the user's problems with a predefined flow. They can switch the conversational scenario when needed and address random user requests at any moment. These chatbots use machine learning, AI, and Natural Language Processing (NLP) to understand and analyse human speech, find the right response and reply in understandable way in a human language.

Overall speaking, chatbots are considered as one of the most advanced and promising aspect of interaction between humans and machines. Chatbot applications helps in smoothening the interactions between the customers and the services [4]. They can enhance and engage customer interactions with less human intervention.

In [5], Dimitrios Buhalis and Iuliia Moldavska explained the importance of voice assistants in the hotel industry. They clearly mention the advantages of voice assistants in hotels outweigh the disadvantages for both hotels and guests. Their findings illustrate that voice-based human-computer interactions bring a range of benefits and voice assistants will be widely deployed in the future. Technology integrations are often complex and costly to set up but it provides significant benefits especially in hotel and tourism industry. As reported by them, guests appreciate the prospective benefits but are concerned with privacy and usability, although tech-savvy consumers are less concerned about privacy when using voice assistants. The findings indicated the direction for the future development of voice technology in hospitality towards multilingualism and modulated offers which can ultimately ensure the overall wider reach of the technology in the hotel industry.

Li, Bai et.al. [6] reviewed a real-world conversational AI and NLP system for hotel booking. Their architecture design includes a frame-based dialogue management system that calls machine learning models for classification, named entity recognition, and information retrieval subtasks. Their chatbot has been deployed on a commercial scale, handling tens of thousands of hotels searches every day. They have also explained various machine learning models that they used for deployment and explained developing an e-commerce chatbot in the travel industry. Adam et.al. [7] describes the significance of chatbots in various fields. They explain how the use of ADCs (Identity, small talk, and empathy) as a common compliance technique, affect user compliance with a request for service feedback in a chatbot interaction. They have examined a randomized online experiment how verbal anthropomorphic design cues and the foot-in-the-door technique affect user request compliance. Their results show that both anthropomorphism as well as the need to stay consistent significantly increase the likelihood that users comply with a chatbot's request for service feedback. They have commented that social presence mediates the effect of anthropomorphic design cues on user compliance. In [8], Hasan et.al. examine tourist chatbot usage intentions in service encounters within the context of a future international travel, assuming continued social distancing. Their results show that automation, habit, social presence, and health consciousness all contributed positively to chatbot usage intentions. Some variations were observed as a function of experiencing government-imposed lockdown. The role of social presence and human qualities in chatbots was weakened when controlling for lockdown and during the trip experience.

## 2. PROPOSED SYSTEM

Our proposed system recognizes speech on chatbot which uses NLP for interaction on any closed domain system. It uses latest technology scope forward and developed the application in internet. Users can use voice to interact with the web application instead of searching and navigate the website. Any questions related to the hospitality or the hotel web application (like how to use, or where can I find it), or questions related to the business domain can get answered by the chatbot.

The system is trained to answer any question related to hospitality domain. The system is trained with any data set which has information related to the hotel website. It is implemented in a way so it can be easily retrained with any data set.

The proposed system requires following modules: a hotel web site which can host the voice chat bot. Web applications need to capture voice input and get the voice to convert into text. There are various solutions for this. Following are two most used voice to text conversion for web application.

### 2.1 SPEECH TO TEXT ENGINE

Deep Speech is an open-source voice recognition and speech to text engine, which provide a trained model using Baidu's Deep Speech research paper and the model implementation is under Mozilla Public license. The underlying implementation is using Google's Tensor Flow. It come up with two models the acoustic model and the language model. Acoustic model is an end-to-end deep leaning system, and the language model is used to increase the accuracy of the transcription output which is included as separate model. The language model can be customized based on our domain.

For the implementation, we must download the model first. The .pbmm is the acoustic model which is trained based on American English, in behind the scene it uses tensor flow. Scorer is the language model, which is useful for improving the accuracy of the predicted output. For example, using this, it will find out which word is grammatically right in a particular context.

The architecture of the engine was originally based up on Deep Speech: Scaling up end-to-end speech recognition. Currently it is different in many aspects and made it based on recurrent neural network (RNN) which is trained to ingest speech spectrogram and generate English text transcription [9]. Deep Speech model use hybrid model for parallel optimization. Hybrid parallel optimization combines the benefit of asynchronous and synchronous optimization. It allows to use multiple GPUs but doesn't have a problem of incorrect gradient present in asynchronous optimization.

In hybrid parallel optimization initially, it places the model in CPU memory. Then, as in asynchronous optimization, each of the G GPUs obtains a mini batch of data along with the current model parameters. Using the mini batch each of the GPUs then computes the gradients for all model parameters and sends these gradients back to the CPU. Now, in contrast to asynchronous optimization, the CPU waits until each GPU is finished with its mini batch then takes the mean of all the gradients from the G GPUs and updates the model with this mean gradient.

Hybrid parallel optimization has several advantages and few disadvantages. As in asynchronous parallel optimization, hybrid parallel optimization allows for one to use multiple GPUs in parallel. Furthermore, unlike asynchronous parallel optimization, the incorrect gradient problem is not present here. In fact, hybrid parallel optimization performs as if one is working with a single mini-batch which is $G$ times the size of a mini-batch handled by a single GPU. However, hybrid parallel optimization is not perfect. If one GPU is slower than all the others in completing its mini-batch, all other GPUs will have to sit idle until this straggler finishes with its mini-batch. This hurts throughput. But, if all GPUs are of the same make and model, this problem should be minimized.

So, relatively speaking, hybrid parallel optimization seems the have more advantages and fewer disadvantages as compared to both asynchronous and synchronous optimization. For this report, we use the hybrid model.

## 2.2 SPEECH RECOGNITION API

In recent years, chrome version 25 came up with the Web Speech API embedded in browser which support conversion of voice to text conversion in web applications. It is getting popular and going to be the future for voice recognition [10]. Browser exposes the speech recognition feature via the Speech Recognition interface. This interface has an ability to recognize voice context from an audio input (normally via the device's default speech recognition service) and respond appropriately. We have to create Speech Recognition object using JavaScript that has a number of event handlers available for detecting when speech is input through the device's microphone. We can also check the browser's compatibility using Webkit Speech Recognition present in browser window object. The Speech Grammar interface represents a container for a particular set of grammar that our app should recognize [11]. Grammar is defined using JSpeech Grammar Format.

## 2.3 TEXT TO VOICE CONVERSION

The output produced by the server will be in text format which need to be converted into voice. So, we need a solution to convert this text into voice in an efficient way. The voice response will provide an interactive feeling to the users. There are many options available in text to voice conversion, for example, many cloud hosted APIs, or standalone implementations using pre-trained models such as gTTS. Here we analysed two solutions such as gTTS and Web Speech API.

### 2.3.1 TEXT TO SPEECH ENGINE USING PRE TRAINED MODEL.

The text-to-speech (TTS) is the process of converting text data into a vocal audio form. The program takes an input text and using methods of natural language processing understands the linguistics of the language being used, and performs logical inference on the text. This processed text is passed into the next block where digital signal processing is performed on the processed text. Using many algorithms and transformations this processed text is finally converted into a speech format. This entire process involves the synthesizing of speech. We use Google's Text To Speech (gTTS) library for text to speech conversion. gTTS is a very easy to use python library which convert text to audio file, which will be transferred to the client as blob data. This API support many languages include English, French, German, Hindi and many more.

**2.3.2 SPEECH SYNTHESIS API**

Speech synthesis is an interface coming up with the browser Web Speech library API. Speech synthesis is accessed via the Speech Synthesis interface, a text-to-speech component that allows programs to read out their text content. It comes up various voice types, rate and pitch that we can configure in the synthesis voice [11].

**2.4 CLOSED DOMAIN QUESTION ANSWERING (cdQA)**

Closed Domain Question Answering (cdQA) is an NLP based Closed Domain Question Answering System. The mission of cdQA is to allow anyone to ask a question in natural language and get an answer without having to read the internal documents relevant to the question.

When we think about question answering systems, it appears as two different kinds of systems: open-domain QA (ODQA) systems and closed-domain QA (cdQA) systems. Open-domain systems deal with questions about nearly anything and can only rely on general ontologies and world knowledge. One example of such a system is DrQA, an ODQA developed by Facebook Research that uses a large base of articles from Wikipedia as its source of knowledge. As these documents are related to several different topics and subjects, we can understand why this system is considered an ODQA. Closed-domain systems deal with questions under a specific domain (for example, medicine or hospitality), and can exploit domain-specific knowledge by using a model that is fitted to a unique-domain database. The cdQA-suite was built to enable anyone who wants to build a closed-domain QA system easily.

The cdQA architecture is based on two main components: the Retriever and the Reader. When a question is sent to the system, the Retriever selects a list of documents in the database that are most likely to contain the answer. It is based on the same retriever of DrQA, which creates TF-IDF (term frequency-inverse document frequency) features based on uni-grams and bi-grams and compute the cosine similarity between the question sentence and each document of the database [11].

After selecting the most probable documents, the system divides each document into paragraphs and send them with the question to the Reader, which is basically a pre-trained Deep Learning model. The model used was the Pytorch version of the well-known NLP model BERT which was made available by HuggingFace. Then, the Reader outputs the most probable answer it can find in each paragraph. After the Reader, there is a final layer in the system that compares the answers by using an internal score function and outputs the most likely one according to the scores.

Here the pretrained model contains Bert Stanford Question Answering Dataset (SQuAD) which have 100,000+ question-answer pairs on 500+ articles. We can further train this model based on our domain and make it a closed cdQA system.

Among the above-mentioned different solutions, by considering various aspects such as accuracy, ease of use, maintainability and cost, the proposed system will be implemented with solution below. We use apache server with PHP as server side scripting language. Voice to Text using Speech recognition API. For text to voice conversion we use Speech Synthesis API. The question answering model use cdQA. The cdQA was hosted on Python Flask web server. The proposed system Architecture diagram is shown in Figure 1.

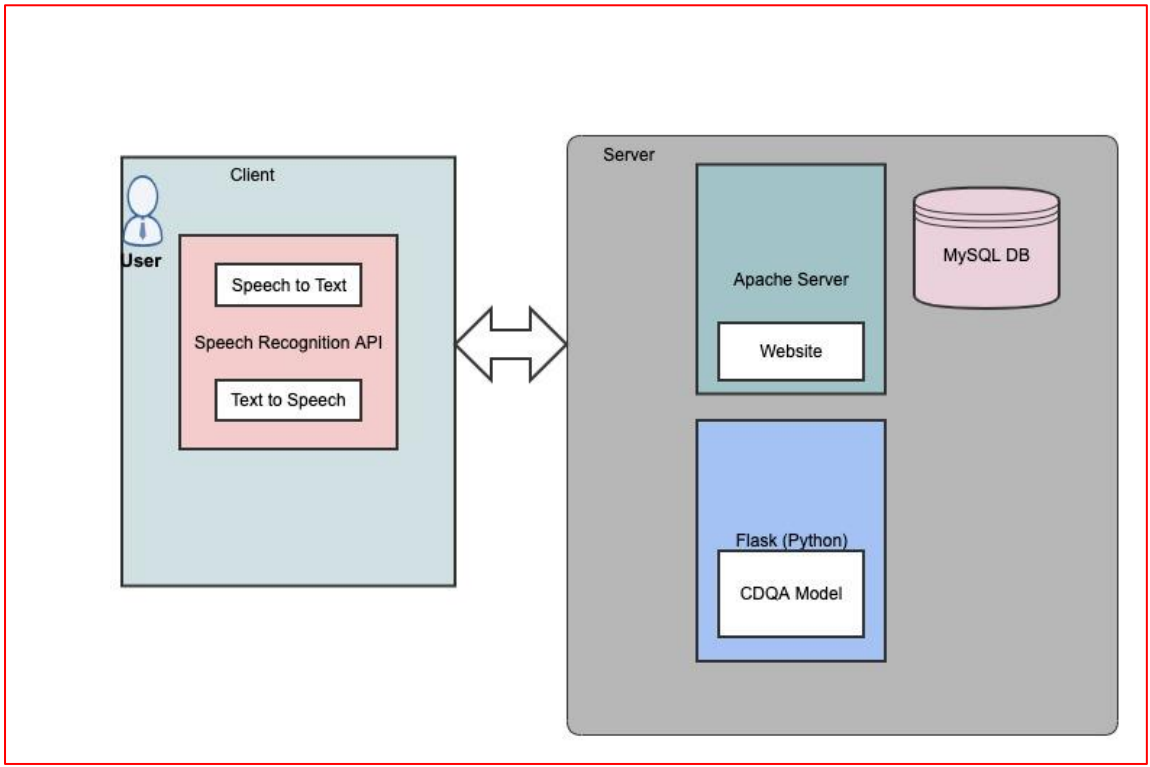

**Figure 1: Proposed System Architecture**

## 3. IMPLEMENTATION AND EXPERIMENTAL RESULTS

Based on the proposed system architecture the system has been implemented with the following. There are five views in this application. The UI shows main attractions nearby, major shopping areas, activities and details about the various kinds of rooms available. The website has been implemented using html, CSS, JavaScript, Bootstrap, FontAwesome for the client side, PHP and apache server for the server side and MySQL database as backend. Room availability is configured in database, it renders in UI when a user search for the rooms. At the top of the website, provided a button to initiate the voice chat. Home screen has been implemented as below, provided search feature at the left, where user can enter the date and guest number to search for the rooms. Click on search button will make rest API call to the server and load the available rooms from the database.

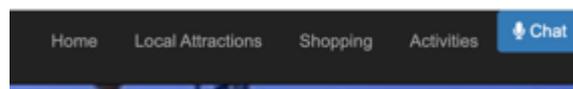

**Figure 2: Home Page Menu**

### 3.1. CLOSED DOMAIN QUESTION ANSWERING (cdQA)

We retrained the model with hospitality pdf documents. Sample response for the question will show as below. The system not only shows outputs an answer, but also shows the paragraph where the answer was found and the title of the document / article.

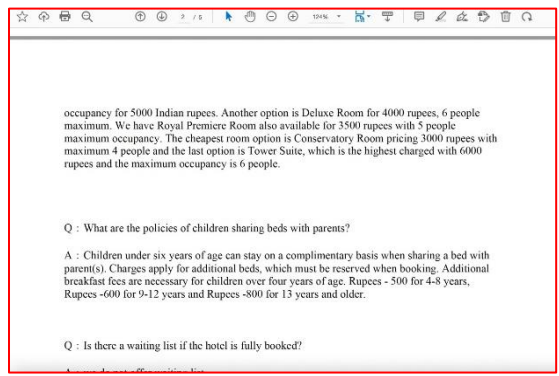

Figure 3: Sample cdQA model pdf

The model has been trained with 50-100 questions related to the hospitality. Each question-answer will be a paragraph. Once the model is trained, deployed, and exposed using Flask server. It exposes an endpoint which accept a string input and query the model and return JSON data which include answer to the query, paragraph where it found the answer and document name. Figure 3 shows a sample cdQA model pdf document.

**3.2 VOICE CHAT BOT (EMMA)**

The designed voice chat bot is named as "Emma". Click on the "Chat (voice)" button at the top the navigation panel will show up a chat box as below with a welcome voice message as "My name is Emma, your voice assistance, how can I help you today". This message is a configurable message.

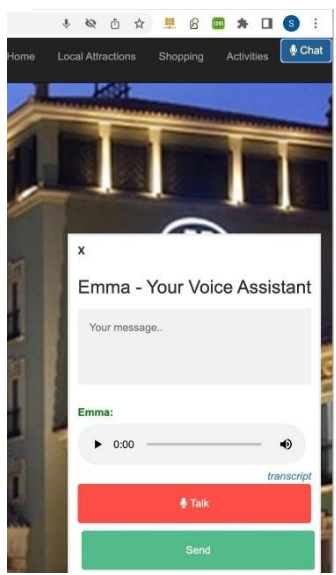

Figure 4: Voice chat bot screen

When user clicks on "Talk" button, it will start listening of voice input till the user stop talking. Using the Web Recognition API, it will convert the input voice into text and will show up in the text area which appears at the top of the voice chat popup as in Figure 5. Click on "send" will initiate a cdQA model hosted server call. The processing status will show up as the loading icon on top left of the Talk button in Figure 5. Note the user's voice was transcribed into text and showed on the screen as well.

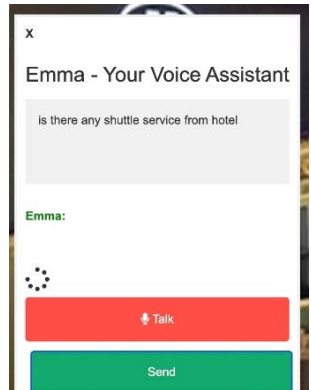

**Figure 5: After user's voice input and the Send button clicked**

Once the result is received, the audio response will render below the text area as shown in Figure 6. It also provides a transcript button, click on that will expand and show the text content.

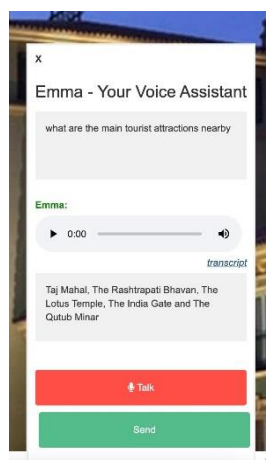

**Figure 6: The voice and text response of the Chat bot**

## 4. CONCLUSIONS AND FUTURE IMPROVEMENTS

Chatbots are software applications that use artificial intelligence and natural language processing to understand humans' need and guides them to their desired outcome. Here this project focused on implementing a voice based chatbot. The voice questions are parsed through an algorithm on a remote server that analyses the document for all possible relevant answers. The most relevant answer is sent back to the user, together with approximate confidence from the model. Reusable generic module which captures voice input and convert it into text and convert the text back to voice once the result received from the server. It is implemented using html and JavaScript library.

Based on the research on various solutions to implement voice to text and text conversion like Google Text-to-Speech (gTTS) engine and speech to text engine, the downside is both require separate infrastructure to maintain it. So decided to proceed with JavaScript utility wrap the implementation for speech recognition and speech synthesis API with event handling.

The question answering mechanism performed very well (based on the evaluation) and predictable for texts shorter than 3000 words. However, with longer texts it started to lose accuracy, losing track of details, and making significant mistakes. The summaries that were meant to help the user formulate questions also worked as intended, however, with the caveat that the summarization model exhibited unaccountable behaviour when supplied with longer texts. The actual usefulness provided are vary between documents. How to conquer this problem may require further research.

An incremental improvement on the system performance and accuracy can be achieved by retraining the model with more questionnaire and through a feedback loop. Training with the bigger dataset requires a higher end system with more GPUs.